\documentclass[reprint,amsmath,amssymb,twocolumn]{revtex4-1}
\usepackage{blindtext}
\usepackage{comment}
\usepackage{subfigure}
\usepackage{graphicx}
\usepackage{dcolumn}
\usepackage{bm}
\usepackage{amssymb}
\usepackage{float}
\usepackage{graphicx}
\usepackage{mathtools}
\usepackage{type1cm}
\usepackage{eso-pic}
\usepackage{color}
\usepackage[scientific-notation=true]{siunitx}
\begin{document}
\preprint{APS/123-QED}

\begin{abstract}
We propose a unifying rheological framework for dense suspensions of non-Brownian spheres, predicting the onsets of particle friction and particle inertia as distinct shear thickening mechanisms, while capturing quasistatic and soft particle rheology at high volume fractions and shear rates respectively.
Discrete element method simulations that take suitable account of hydrodynamic and particle-contact interactions corroborate the model predictions, demonstrating both mechanisms of shear thickening, and showing that they can occur concurrently with carefully selected particle surface properties under certain flow conditions.
Microstructural transitions associated with frictional shear thickening are presented. We find very distinctive divergences of both the microstructural and dynamic variables with respect to volume fraction in the thickened and non-thickened states.
\end{abstract}

\title{Shear thickening regimes of dense non-Brownian suspensions}
\author{Christopher Ness}
\email{christopher.ness@cantab.net}
\affiliation{School of Engineering, University of Edinburgh, Edinburgh EH9 3JL, United Kingdom}%
\author{Jin Sun}
\email{j.sun@ed.ac.uk}
\affiliation{School of Engineering, University of Edinburgh, Edinburgh EH9 3JL, United Kingdom}%

\date{\today}
\maketitle

\section{Introduction}

Non-Newtonian rheology~\cite{Denn2004} has been observed and studied for centuries in numerous materials, flow regimes and applications. In this work we focus on shear thickening~\cite{Brown2013} in densely packed non-Brownian suspensions of bidisperse solid spheres, with and without inertia~\cite{Stickel2005,Lemaitre2009,DeBruyn2011}. 
This rheological phenomenon, in which the shear stress required to deform the suspension increases faster than linearly with the deformation rate, is regularly demonstrated in high volume fraction cornstarch suspensions~\cite{Fall2008}, but is also observed in other particulate systems such as dry granular materials at constant volume~\cite{Chialvo2012,Forterre2008,Jop2006} and well characterised model suspensions~\cite{Brown2009}, and has considerable industrial relevance~\cite{Benbow1993}. The non-Brownian limit arises in suspensions of both silica and polymethylmethacrylate, for example, under typical shear thickening conditions~\cite{Lin2015a}.

Continuous, linear shear thickening, in which the suspension viscosity is proportional to the shear rate, may arise in suspensions below jamming~\cite{Liu1998} when conditions are such that particle inertia is relevant~\cite{Fall2010,Trulsson2012,Ness2015}, much like in dry granular materials~\cite{Midi2004}. Other suspensions have, however, been observed to shear thicken far more severely than in these linear cases, and at Stokes numbers considerably less than 1, for which particle inertia ought to be negligible~\cite{Fall2008,Fall2012,Fall2015,Laun1984,Egres2005,Cwalina2014}. This behaviour is variously known as ``shear jamming", ``dynamic jamming" and ``discontinuous shear thickening'' (DST)~\cite{Bertrand2002}, and until recently has widely been thought to arise due to either the shear-induced formation of ``hydroclusters''~\cite{Wagner2009,Cheng2011}, mesoscale particle agglomerates stabilised by hydrodynamic interactions that result in massive dissipation under shear; or dilatancy, the tendency of the suspension to increase in volume upon shearing~\cite{Brown2012a, Fall2008} and subsequently bifurcate into coexisting regimes of inhomogeneous solids fraction~\cite{Fall2015}.

A growing body of experimental~\cite{Fernandez2013,pan2015} and computational~\cite{Seto2013,Mari2014a,Heussinger2013} work provides evidence that discontinuous shear thickening can arise because frictional particle-particle contacts appear under large loads. Such suspended particles may be either charge stabilised or sterically stabilised using, for example, polymer hairs grafted to the particle surface. Under small loads, the normal repulsive forces that arise between particles due to this stabilisation are sufficient to prevent direct particle-particle contacts, so lubricating layers are maintained. Above a critical load $P^*$, the stabilisation is overcome and rough particle surfaces come into contact, resulting in normal and tangential forces that can be considered similar to those existing between dry granular particles~\cite{Guy2015a}. The increased dissipation resulting from the subsequently reorganised microstructure and the tangential contact forces means very large stresses are required to maintain flow. Under this mechanism, the shear thickened state may flow homogeneously, without velocity or volume fraction banding~\cite{pan2015}.
A rheological model proposed by Wyart and Cates~\cite{Wyart2014} captures this transition between frictionless and frictional states, predicting the presence of continuous shear thickening at low volume fractions, and DST at high volume fractions, where S-shaped flow curves could occur with multiple flow states existing at a given shear rate but at greatly differing stresses. Such flow curves have recently been observed experimentally~\cite{pan2015} and computationally~\cite{Mari2015} under imposed shear stress.

In general, the rheology of suspended ``nearly-hard'' spheres can be broadly characterised by the interplay between the flow timescale associated with inverse of the shear rate $1/\dot{\gamma}$, and \emph{four} competing timescales: a Brownian timescale (characterised by the P\'{e}clet number $\mathrm{Pe} = 3\pi \eta_f \dot{\gamma}d^3/4kT$, for representative particle diameter $d$ and interstitial fluid viscosity $\eta_f$); a timescale associated with the stabilising repulsion (numerically this has been referred to as $1/\dot{\gamma}_0 = \frac{3}{2}\pi\eta_f d^2/F^{CL}$, for repulsive force magnitude $F^{CL}$)~\cite{Mari2014a}; an inertial timescale (characterised by the Stokes number $\mathrm{St} = \rho d^2 \dot{\gamma}/\eta_f$, where $\rho$ is a representative suspension density and $\mathrm{St}=1$ delineates viscous and inertial flows) and a timescale associated with the stiffness of the particles (for example $d/\sqrt{k_n/\rho d}$, where $k_n$ is related to the Young's modulus of the particles)\cite{Chialvo2012}. So far, these diverse scales have not been probed simultaneously in a single suspension either experimentally or computationally, though there are numerous recent examples of transitions across regimes that suggest they represent isolated regions of a single rheological map for dense suspensions~\cite{Mari2015a, Ikeda2012, Kawasaki2014b,Cwalina2014,Guy2015a,Nordstrom2010a, Fall2010}.

In the present work, we focus on the non-Brownian limit (i.e. $\text{Pe}\to\infty$), and demonstrate that constitutive models proposed for shear thickening in the non-inertial limit~\cite{Wyart2014} and for capturing the non-inertial (viscous) to inertial transition~\cite{Trulsson2012,Ness2015}, can be unified to place frictional shear thickening in the wider context of dense suspension rheological regimes, also accounting for the effects of finite particle hardness.
We then perform discrete element method~\cite{Cundall1979,Plimpton1995} simulations combining hydrodynamic lubrication~\cite{Ball1997} with a suitable particle-particle interaction model proposed by Mari et al.~\cite{Seto2013,Mari2014a} that can capture the bulk steady-state rheological behaviour associated with frictional shear thickening under imposed shear rate. We demonstrate that the timescale for \emph{frictional} shear thickening can be made to coincide with that for \emph{inertial} shear thickening by careful tuning of particle surface properties, hinting at novel suspension flow curves that have yet to be observed experimentally.
Finally, we highlight microstructural properties associated with the frictional thickening transition, identifying very well defined structural and dynamic signatures that may prove useful in interpretation and analysis of future rheo-imaging data for shear thickening suspensions.

\section{Constitutive modelling and flow regime map}
\label{sec:flow_map}

\begin{figure}
\includegraphics[width=80mm]{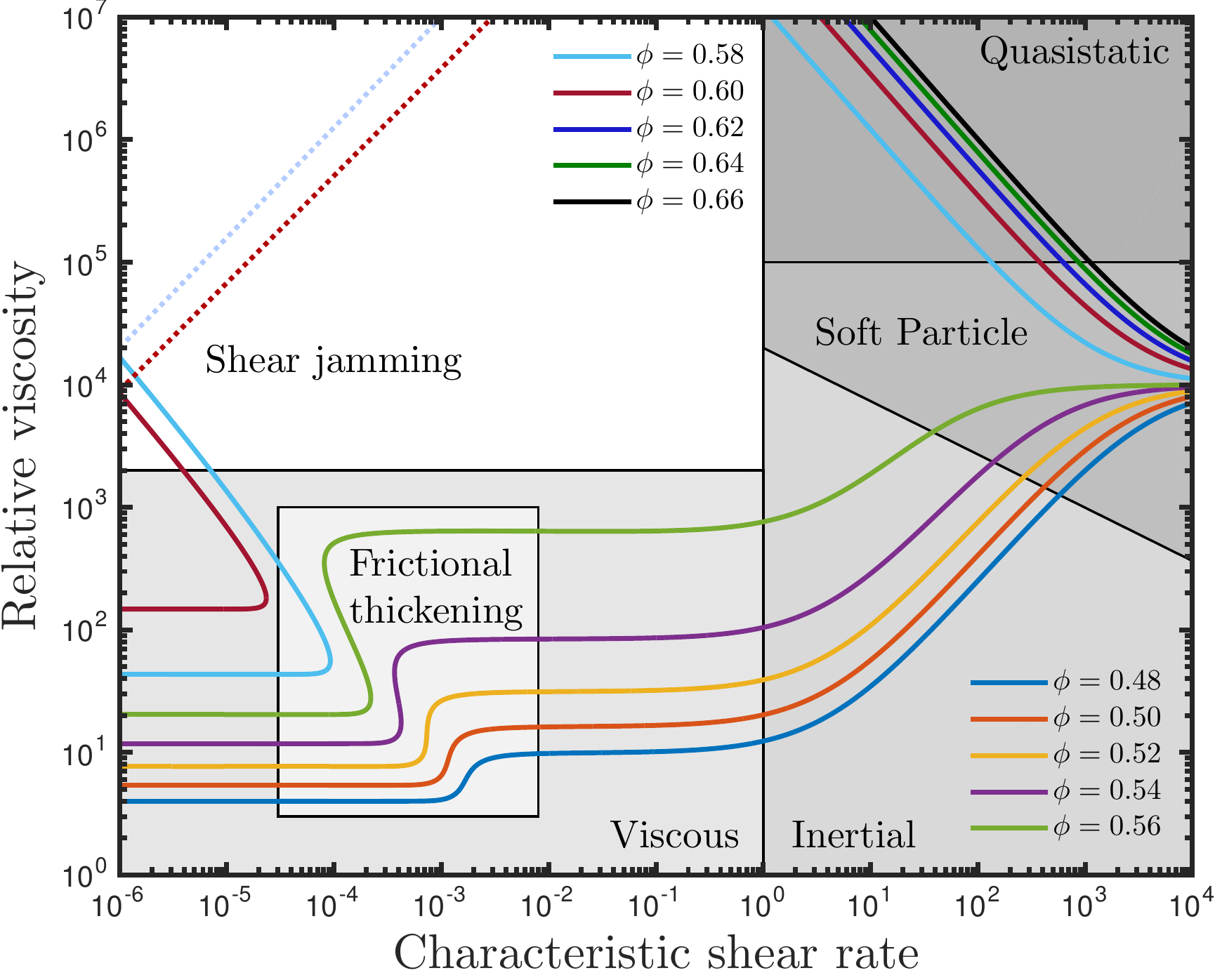}
 \caption{Steady state rheological regime map for a shear thickening suspension, illustrating the frictional thickening transition within the viscous regime, shear jamming, inertial shear thickening, quasistatic behaviour and deformational behaviour associated with soft particle rheology.}
 \label{fig:regime_map}
\end{figure}

We first present a rheological equation that is able to capture the viscous, inertial, quasistatic and soft-particle flow regimes~\cite{Ness2015}. This model is inspired by the inertial number model~\cite{Jop2006,Forterre2008} (for inertial number $I_I=\dot{\gamma}d/\sqrt{P/\rho}$, with confining pressure $P$), its extension to viscous flows~\cite{Boyer2011} (for viscous number $I_V=\dot{\gamma}\eta_f/P$), and their proposed unification by Trulsson~\cite{Trulsson2012}. The equation gives a prediction for the scaled (by particle hardness) pressure ($\hat{P} = Pd/k_n$) as a function of the scaled shear rate ($\hat{\dot{\gamma}} = \dot{\gamma}d/\sqrt{k_n/\rho d}$) and the departure of the solids volume fraction $\phi$ from its critical value for jamming~\cite{Liu1998}, $\phi_c$, in each of three regimes: 1) the hard particle regime corresponding to viscous and inertial flows; 2) the soft particle regime corresponding to deformable particle flows; 3) the quasistatic, ``jammed'' regime:
\begin{subequations}
\begin{equation}
\hat{\text{P}}_\text{hard} = 
\underbrace{\alpha_\text{hard}^{\text{c}}|\phi - \phi_c|^{-2} \hat{\dot{\gamma}}^{2}}_\text{contact} 
+
\underbrace{\alpha_\text{hard}^{\text{f}} {\eta}_f|\phi - \phi_c|^{-2} \hat{\dot{\gamma}}}_\text{fluid} \text{,}
\end{equation}
\begin{equation}
\hat{\text{P}}_\text{soft} = 
\underbrace{\alpha_\text{soft}^{\text{c}} \hat{\dot{\gamma}}^{0.5}}_\text{contact}
+
\underbrace{\alpha_\text{soft}^{\text{f}} \hat{\eta}_f \hat{\dot{\gamma}}}_\text{fluid}\text{,}
\end{equation}
\begin{equation}
\hat{\text{P}}_\text{QS} = 
\underbrace{\alpha_\text{QS}|\phi - \phi_c|^{2/3}}_\text{contact}\text{,}
\end{equation}
\label{eq:ness2015}
\end{subequations}
with the constants given by Ness and Sun~\cite{Ness2015}. An arbitrary blending function is chosen, following Chialvo et al.~\cite{Chialvo2012}, to combine pressure predictions from each of the expressions.
The corresponding shear stresses $\sigma_{xy}$ are obtained from a $\mu(K=I_V+\alpha I_I^2)$ model~\cite{Trulsson2012,Ness2015}, an extension of the commonplace $\mu(I_I)$ rheology~\cite{Jop2006}:
\begin{equation}
\mu(K) = \mu_1 + 1.2 K ^ {1/2} + 0.5 K \text{.}
\label{eq:mu}
\end{equation}
The rheology predicted by Equations~\ref{eq:ness2015}~and~\ref{eq:mu} is presented in detail in Ref.~\cite{Ness2015}.
We next incorporate a frictional shear thickening mechanism into this model using a stress-dependent critical volume fraction $\phi_c$, following the approach used by Wyart and Cates~\cite{Wyart2014}. A stress-dependent effective friction is introduced, recognising that the critical volume fraction depends on the interparticle friction coefficient $\mu_p$~\cite{Sun2011}, varying from $\phi_{m} \approx 0.58$ to $\phi_{0} \approx 0.64$ in the limits of frictional ($\mu_p=1$) and frictionless ($\mu_p=0$) particles, respectively.
From the simulation model described in Sec~\ref{sec:methods}, we find that under shear flow, the rescaled pairwise particle-particle contact force magnitudes $\theta = |\mathbf{F}^c_{ij}|/Pd^2$ are distributed according to
\begin{equation}
\text{PDF}(\theta) = a(1-b \exp(-\theta^2))\exp(-c \theta) \text{,}
\end{equation}
consistent with previous authors\cite{Mueth1998,Sun2006}.
The fraction of particle contacts for which the repulsive force magnitude $F^{CL}$ is exceed and friction is activated is therefore given by
\begin{equation}
f = \int_{F^{CL}/Pd^2}^\infty \text{PDF}(\theta)d\theta \text{,}
\end{equation}
implying (except for very weak contacts) that frictional forces arise in the system above $P^*$ according to $f\propto\exp(-P^*/P)$. We therefore use $f$ to represent a transition from frictionless to frictional rheology with increasing $P$.
The value set for $P^*$ (which is directly related to the repulsive force magnitude $F^{CL}$) determines the critical pressure (or critical characteristic shear rate) at which the model will begin to predict frictional rheology, as described later. We subsequently calculate the (stress-dependent) critical volume fraction for jamming $\phi_c$ using an expression similar to the crossover function proposed by Wyart and Cates\cite{Wyart2014}
\begin{equation}
\phi_c = \phi_{m} f + \phi_{0}(1-f)\text{.}
\label{eq:phi_c}
\end{equation}
The expression for $\mu(K)$, along with that proposed by~\cite{Boyer2011}, assumes a constant value for the macroscopic friction $\mu_1$ ($=\sigma_{xy}/P = 0.38$) in the limit of $K\to 0$. 
As demonstrated by da Cruz~\cite{daCruz2005}, $\mu_1$ is actually strongly dependent on interparticle friction $\mu_p$, particularly for $\mu_p$ close to 0, so assuming a constant value across shear thickening states is clearly not appropriate. We therefore propose a similar crossover function for $\mu_1$ (consistent with that proposed by Sun and Sundaresan~\cite{Sun2011} for dry granular materials) which we find gives excellent agreement with the following simulation results
\begin{equation}
\mu_1 = \mu_{1_{m}}f+ \mu_{1_0}(1-f) \text{,}
\label{eq:mu_1}
\end{equation}
where $\mu_{1_{m}} = 0.41$ and $\mu_{1_{0}} = 0.11$~\cite{Sun2011}. We obtain the viscosity of the suspension relative to that of the interstitial fluid according to $\eta_s = \frac{\sigma_{xy}}{\eta_f \dot{\gamma}}$.

We obtain the flow curves presented in Fig~\ref{fig:regime_map} from Eqs.~\ref{eq:ness2015}, \ref{eq:mu}, \ref{eq:phi_c} and \ref{eq:mu_1}. Below $\phi_c$, the model predicts viscous rheology for Stokes numbers less than unity, inertial flow at higher Stokes numbers, and ``intermediate'' rheology at extremely high shear rates or for soft particle suspensions (i.e. emulsions) where large deformations are possible~\cite{Nordstrom2010a} (strictly, ``soft'' particle rheology is observed when $\dot{\gamma}\to d\sqrt{k_n/\rho d}$)~\cite{Chialvo2012}. Above $\phi_c$, quasistatic rheology is observed for low and moderate Stokes numbers, with a tendency towards soft particle rheology at very high rates.
The addition of pressure dependence in $\phi_c$ gives rise to the frictional thickening and S-shaped flow behaviour, as shown in Fig~\ref{fig:regime_map} within the viscous flow regime, and the hypothetical shear jamming transition that may occur between the viscous and quasistatic regimes for particles of finite stiffness. The fact that the onset stress $P^*$ varies with particle contact properties (specifically $F^{CL}$, the repulsive force magnitude) implies that the transition to frictional behaviour might occur at different regions of the flow map. We demonstrate this behaviour in Fig~\ref{fig:flow_curves}, by increasing (from Fig~\ref{fig:flow_curves}(a) to (c))  the magnitude of the onset stress $P^*$ in the definition of $f$, delaying the onset of the frictional behaviour governed by Eqs.~\ref{eq:phi_c} and \ref{eq:mu_1} such that it occurs at higher Stokes numbers. We obtain the same characteristic frictional shear thickening flow curve predictions for each of Figs~\ref{fig:flow_curves}a-c, with the added phenomena of inertia appearing at a prescribed point in relation to $P^*$ (in \ref{fig:flow_curves}b~and~\ref{fig:flow_curves}c). Such shear thickening will be demonstrated by particle simulations in the next section.

\section{Shear flow simulations}
\label{sec:steady_rheology}
\subsection{Numerical method}
\label{sec:methods}
The equations of motion for non-Brownian particles suspended in a fluid can be written simply as~\cite{Brady1988}
\begin{equation}
m \frac{d}{dt}\binom{\mathbf{v}}{\boldsymbol{\omega}} = \sum \binom{\mathbf{F}}{\bm{\Gamma}} \text{,}
\end{equation}
for particles of mass $m$ with translational and rotational velocity vectors $\mathbf{v}$ and $\boldsymbol{\omega}$ respectively, subjected to force and torque vectors $\mathbf{F}$ and $\bm{\Gamma}$ respectively. In this work we limit the forces and torques to those arising due to direct particle contacts ($\mathbf{F}^c, \bm{\Gamma}^c$) and those arising through hydrodynamic interactions ($\mathbf{F}^l, \bm{\Gamma}^l$). Full solution of the pairwise hydrodynamic forces has traditionally been done using the Stokesian Dynamics algorithm~\cite{Bossis1984, Brady1985}, though its great computational expense makes large (or very dense) simulations challenging. For highly packed suspensions, the divergent lubrication resistances that arise between extremely close particles dominate the hydrodynamic interaction, so $\mathbf{F}^l$, $\bm{\Gamma}^l$ can be approximated by summing pairwise lubrication forces among nearest neighbouring particles~\cite{Ball1997,Kumar2010, Trulsson2012,Mari2014a,Ness2015}. For an interaction between particles $i$ and $j$, the force and torque due to hydrodynamics can therefore be expressed as 
\begin{subequations}
\begin{align}
&
\begin{multlined}
\mathbf{F}^{l}_{ij} = -a_{sq} 6 \pi \eta_f (\textbf{v}_i - \textbf{v}_j) \cdot \textbf{n}_{ij} \textbf{n}_{ij}\\
- a_{sh} 6 \pi \eta_f (\textbf{v}_i - \textbf{v}_j) \cdot (\mathbf{I}-\mathbf{n}_{ij}\mathbf{n}_{ij}) \text{,}
\end{multlined}\\
&
\begin{multlined}
\bm{\Gamma}^{l}_{ij} = -a_{pu} \pi \eta_f d_i^3(\boldsymbol{\omega}_i - \boldsymbol{\omega}_j) \cdot (\mathbf{I}-\mathbf{n}_{ij}\mathbf{n}_{ij})\\
- \frac{d_i}{2} \left(\mathbf{n}_{ij} \times \mathbf{F}^{l}_{ij}\right) \text{.}
\end{multlined}
\end{align}
\label{eq:lube_forces}
\end{subequations}
where $\mathbf{n}_{ij}$ is the vector pointing from particle $j$ to particle $i$, and with squeeze $a_{sq}$, shear $a_{sh}$ and pump $a_{pu}$ resistance terms as derived by Kim and Karrila\cite{Kim1991} and given in Eq~\ref{eq:resistances} for particle diameters $d_i$ and $d_j$, with $\beta = d_j/d_i$:

\begin{subequations}
\begin{align}
&
\begin{multlined}
a_{sq} = \frac{\beta^2}{(1+\beta)^2} \frac{d_i^2}{2h_\text{eff}}	+	\frac{1 + 7\beta + \beta^2}{5(1 + \beta)^3} \frac{d_i}{2} \ln \left(\frac{d_i}{2h_\text{eff}} \right)\\
+ \frac{1 + 18\beta - 29\beta^2 + 18\beta^3 + \beta^4}{21(1+\beta)^4} \frac{d_i^2}{4h_\text{eff}} \ln \left(\frac{d_i}{2h_\text{eff}}\right) \text{,}
\end{multlined}\\
&
\begin{multlined}
a_{sh} = 4 \beta \frac{2 + \beta + 2\beta^2}{15 (1 + \beta)^3} \frac{d_i}{2} \ln \left( \frac{d_i}{2h_\text{eff}}\right)\\
+ 4\frac{16 -45\beta + 58\beta^2 - 45\beta^3 + 16\beta^4}{375(1 + \beta)^4} \frac{d_i^2}{4h_\text{eff}} \ln \left( \frac{d_i}{2h_\text{eff}} \right) \text{,}
\end{multlined}\\
&
\begin{multlined}
a_{pu} = \beta \frac{4 + \beta}{10(1 + \beta)^2} \ln \left( \frac{d_i}{2h_\text{eff}} \right)\\
+ \frac{32 - 33\beta + 83\beta^2 + 43\beta^3}{250\beta^3} \frac{d_i}{2h_\text{eff}} \ln \left( \frac{d_i}{2h_\text{eff}} \right) \text{.}
\end{multlined}
\end{align}
\label{eq:resistances}
\end{subequations}
For each pairwise interaction, the surface-to-surface distance, $h$, is calculated according to $h = |\mathbf{r}_{ij}| - \frac{d_i + d_j}{2}$, for centre-to-centre vector $\mathbf{r}_{ij}$.
Recent experimental \cite{Fernandez2013,Guy2015a} and computational \cite{Lin2015a,Ness2015a} work indicates that direct particle-particle contacts play a significant role in determining steady state paste viscosity. To permit such contacts in the present model, we truncate the lubrication divergence and regularize the contact singularity at a typical asperity length scale $h_\text{min}=0.001d_{ij}$ (for weighted average particle diameter $d_{ij} = \frac{d_id_j}{d_i + d_j}$), i.e., setting $ h = h_\text{min}$ in the force calculation, when $h < h_\text{min}$.
The effective interparticle gap used in the force calculation, $h_\text{eff}$, is therefore given by
\begin{equation}
    h_\text{eff} =  \left\{
                \begin{array}{ll}
                  h & \text{for } h > h_\text{min}\\
                  h_\text{min} & \text{otherwise.} 
                \end{array}
              \right.
\end{equation}
For computational efficiency, the lubrication forces are omitted when the interparticle gap $h$ is greater than $h_\text{max} = 0.05 d_{ij}$. The volume fraction is sufficiently high that all particles have numerous neighbours within this range, so such an omission is inconsequential to the dynamics.
When the lubrication force is overcome and particle surfaces come into contact (this occurs when $h<0$), their interaction is defined according to a linear spring model~\cite{Cundall1979}, with normal ($\mathbf{F}^{c,n}$ ) and tangential ($\mathbf{F}^{c,t}$ ) force and torque $\bm{\Gamma}^c$ given by
\begin{subequations}
\begin{equation}
\mathbf{F}^{c, n}_{ij} = k_\text{n} \delta \mathbf{n}_\text{ij} \text{,}
\end{equation}
\begin{equation}
\mathbf{F}^{c,t}_{ij} = -k_\text{t} \mathbf{u}_\text{ij} \text{,}
\end{equation}
\begin{equation}
\bm{\Gamma}^{c}_{ij} = -\frac{d_i}{2} (\mathbf{n}_{ij} \times \mathbf{F}^{c,t}_{ij}) \text{,}
\end{equation}
\end{subequations}
for a collision between particles $i$ and $j$ with normal and tangential spring stiffnesses $k_n$ and $k_t$ respectively, particle overlap $\delta$ (equal to $-h$) and tangential displacement $\mathbf{u}_\text{ij}$. We note that the damping arising from the hydrodynamics is always sufficient to achieve a steady state without employing a thermostat, and further damping in the particle-contact model is omitted for simplicity.

We employ the Critical Load Model (CLM) for inter-particle friction~\cite{Seto2013, Mari2014a}, to distinguish between weakly interacting particles, those that interact via the normal force deriving from stabilisation, and strongly interacting particles, those whose surfaces come into frictional contact.
This model gives an additional stress scale for the particle interaction, which, numerically, is the origin of the onset stress for shear thickening $P^*$.
An interparticle Coulomb friction coefficient $\mu_p$ is defined according to  $|\mathbf{F}^{c,t}_{i,j}| \leq \mu_p|\mathbf{F}^{c,n}_{i,j}| $, setting a maximum value for the tangential force exerted during a collision. For each pairwise collision, the value of $\mu_p$ is dependent upon the normal force between the interacting particles and some critical normal force magnitude $F^{CL}$, representing the magnitude of the stabilising repulsive force, such that
\begin{equation}
    \mu_p =  \left\{
                \begin{array}{ll}
                  1 & \text{for } |\mathbf{F}^{c,n}_{i,j}| > F^{CL}\\
                  0 & \text{otherwise} 
                \end{array} \text{.}
              \right.
\end{equation}
As a result of the CLM, particles that interact through weak forces, i.e. collisions where $\delta \to 0$, are frictionless, while interactions with large normal forces are frictional.
This particle potential represents a physical scenario closer to electrostatic rather than polymer hair driven normal repulsion. 
The particle overlaps required to exceed $F^{CL}$ are, at their absolute maximum, of order $10^{-5}d_{ij}$. 

Isotropic particle assemblies are generated in a 3-dimensional periodic domain of volume $V$. It is determined that approximately 5000 spheres are sufficient to capture the bulk rheology and microstructural phenomena independently of the system size. Bidispersity at a diameter ratio of $1:1.4$ and volume ratio of $1:1$ is used to minimize crystallization during flow~\cite{Ikeda2012}.
The particle assembly is sheared to steady state at constant rate $\dot{\gamma}$ and constant volume, equivalent to the application of Lees-Edwards boundary conditions~\cite{Lees1972}.
The bulk stress, decomposed into contributions due to the hydrodynamic interaction and the particle-particle interaction, is calculated from the particle force data~\cite{Ness2015}, and given by Eqs.~\ref{eq:stressF}~and~\ref{eq:stressC}
\begin{subequations}
\begin{align}
\bm{\sigma}^F &= \frac{1}{V} \sum_i \sum_{i \neq j} \mathbf{r}_{ij} \mathbf{F}^l_{ij} \text{,} 
\label{eq:stressF} \\
\bm{\sigma}^C &= \frac{1}{V} \sum_i \sum_{i \neq j} \mathbf{r}_{ij} \mathbf{F}^c_{ij} \text{.}
\label{eq:stressC}
\end{align}
\end{subequations}
Data from 20 realizations with randomized initial particle positions are used to obtain ensemble-averaged stresses, which are further averaged over time in the steady-state. Under simple shear flow, the relevant stresses that will be discussed are the $xy$ component $\sigma_{xy}$ ($=\sigma^F_{xy} + \sigma^C_{xy}$), for flow direction $x$ and gradient direction $y$, and the mean normal stress $\text{P} = \frac{1}{3} (\sigma_{xx} + \sigma_{yy} + \sigma_{zz})$, for $\sigma_{xx} = \sigma^F_{xx} + \sigma^C_{xx}$ etc.

\begin{figure}
  \centering
      \subfigure[]{\includegraphics[width=75mm]{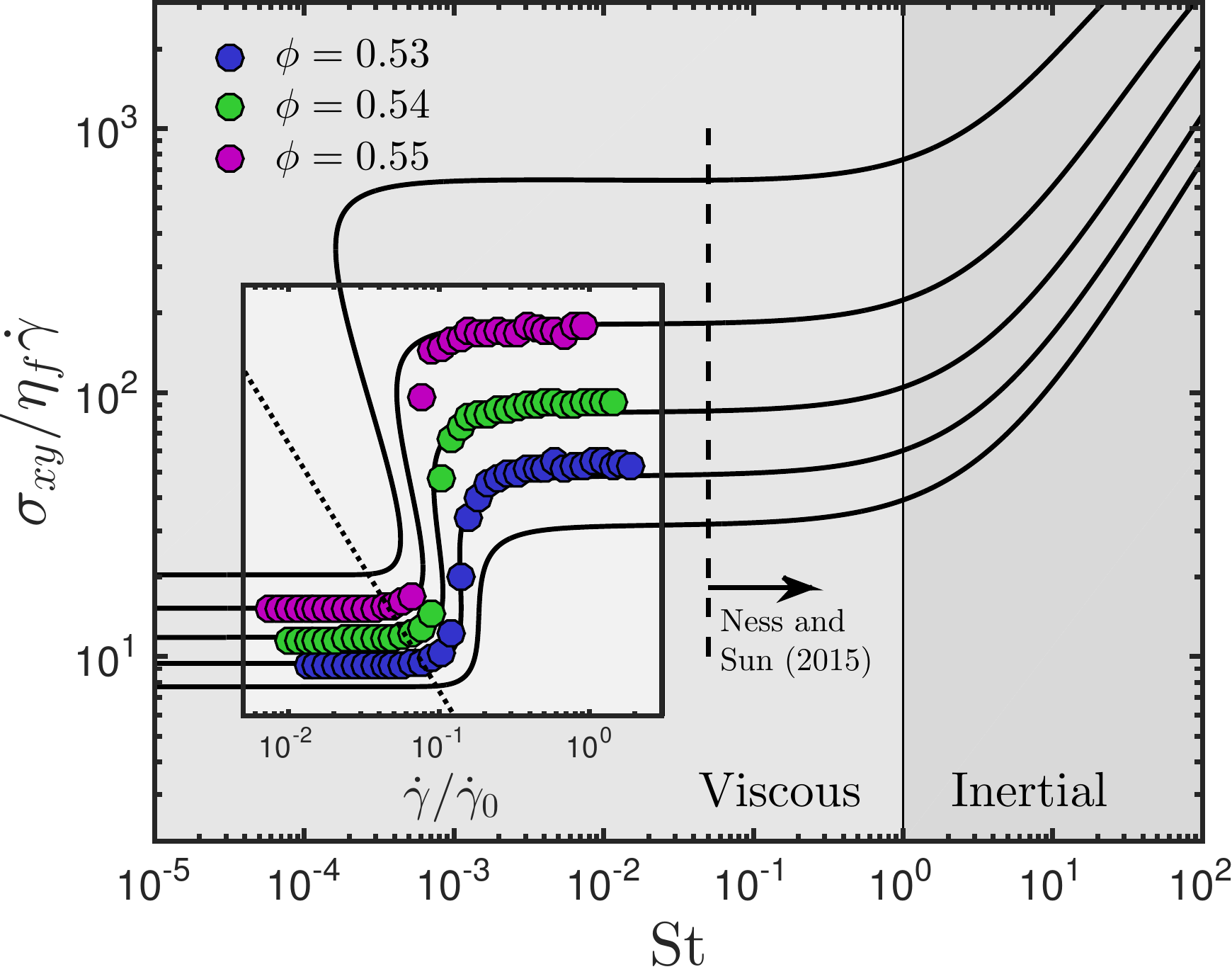}}
      \subfigure[]{\includegraphics[width=75mm]{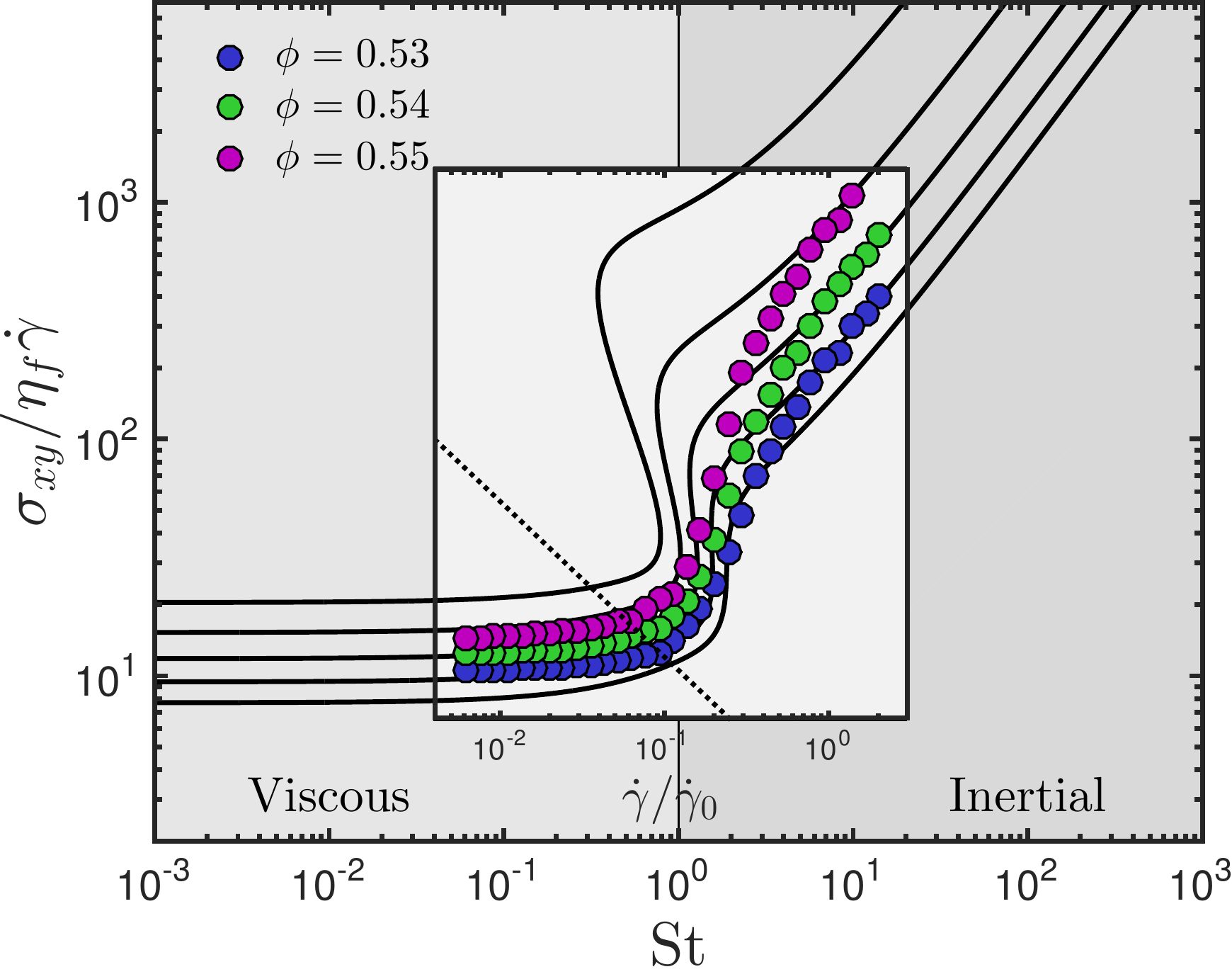}}
      \subfigure[]{\includegraphics[width=75mm]{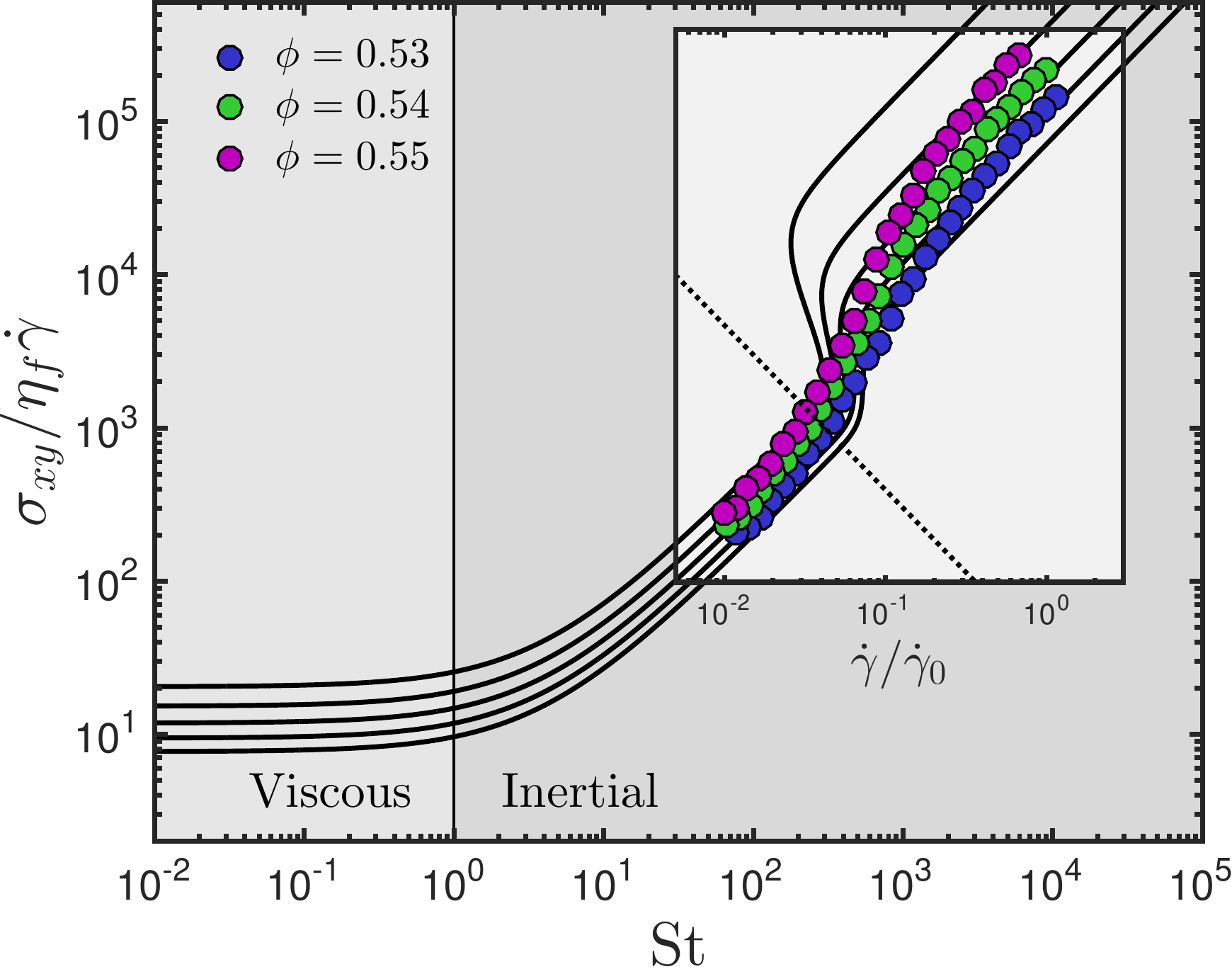}}
     \caption{
Shear thickening transition for three values of $P^*$ (or, equivalently, $F^{CL}$).
(a) Frictional shear thickening occurs in the absence of inertia. Dashed line and arrow demonstrates the relative location of rheological data presented by Ness and Sun~\cite{Ness2015};
(b) Frictional shear thickening occurs concurrently with the onset of inertia;
(c) Frictional shear thickening occurs in the presence of inertia.
Coloured circles represent discrete element method simulation results; solid black lines represent constitutive model predictions; dotted black lines represent $P^*$.
 }
 \label{fig:flow_curves}
\end{figure}


\subsection{Macroscopic flow behaviour}

Transitions between the viscous, inertial, soft particle and quasistatic regimes, as they are depicted in Fig~\ref{fig:regime_map}, have been previously captured by discrete element method simulations and well characterised~\cite{Ness2015}. 
The steady state shear thickening behaviour predicted by the simulation model described in Sec~\ref{sec:methods} is presented as solid coloured symbols in Fig~\ref{fig:flow_curves}. We first focus on the results in Fig~\ref{fig:flow_curves}a, which correspond directly to the frictional thickening transition highlighted within the viscous flow regime in Fig~\ref{fig:regime_map}.
Following~\cite{Mari2014a}, the shear rate $\dot{\gamma}$ is scaled with the reciprocal of a characteristic timescale for the relaxation of a frictional contact in a viscous fluid, given by $\dot{\gamma}_0 = F^{CL} / \frac{3}{2}\pi \eta_f d^2$. 
Consistent with the results obtained by Mari et al.~\cite{Seto2013,Mari2014a}, shear thickening between two quasi-Newtonian flow regimes is observed to occur at an onset stress $P^*$, independent of volume fraction and given by the dashed black line in Fig~\ref{fig:flow_curves}a. 
Far below the onset stress, particles interact through forces predominantly $|\mathbf{F}^{c,n}_{i,j}|<F^{CL}$, that is, the forces are not sufficiently large to overcome the stabilisation, so frictional particle surfaces do not often come into contact. Conversely, the stabilisation is nearly always overcome (so contacts are nearly always frictional) at stresses far above the onset stress. We therefore make a distinction between purely frictionless behaviour at $\dot{\gamma}/\dot{\gamma}_0 = 0.01$ and purely frictional behaviour at $\dot{\gamma}/\dot{\gamma}_0 = 1$.
The solid black lines represent predictions from the constitutive model described previously. The value of the onset stress is determined by the magnitude of $F^{CL}$ specified in the contact potential, and is inferred from the simulation data. The annotation in Fig~\ref{fig:flow_curves}a illustrates the relative position of the simulation data presented by Ness and Sun~\cite{Ness2015}.

For volume fractions below approximately $\phi =0.53$, the rheology exhibits continuous shear thickening behaviour, while between $\approx$0.53 and $\approx$0.58 the thickening is discontinuous, in that the constitutive model flow curves (solid black lines) exhibit the characteristic `S-shaped' phenomena. The simulation data do not populate the `S-shaped' region, probably because the simulations were performed for steady states at enforced constant shear rate, while the nature of flow in this regime is high unstable in reality. Probing the `S-shaped' region through other simulation protocols is the subject of ongoing investigation and will be reported on in the future. 
For volume fractions above $\phi_m$, the material ``shear-jams'' above $P^*$, as illustrated in Fig~\ref{fig:regime_map}. When the onset stress is exceeded above $\phi_m$, the material transitions from below to above its critical volume fraction, meaning the flow moves from a flowing, viscous state to a jammed state. Experimentally, this may be manifested as complete flow cessation, surface fracture, microstructural inhomogeneity, or volume fraction bifurcation, depending on the nature of the rheometer. Particle overlaps are allowed in the simulations, so the flow can enter a quasistatic state above jamming, in which the shear stress is rate-independent~\cite{Ness2015}. 



To bring the frictional thickening transition nearer to the inertial regime computationally, we simply reduce the viscosity of the interstitial fluid $\eta_f$, modifying $\dot{\gamma}_0 (=F^{CL}/\frac{3}{2}\pi\eta_f d^2_{ij}$) and effectively moving the $0.01<\dot{\gamma}/\dot{\gamma}_0<0.1$ window to a higher range of Stokes numbers. We can achieve an analogous effect by adjusting $F^{CL}$, comparable to modifying either the particle size or the particle surface chemistry experimentally.
In terms of shear thickening, the effect of this adjustment is to alter the magnitude of the onset stress for frictional contacts such that it occurs in the vicinity of any desired Stokes number. Flow curves are presented for an onset stress that occurs close to St$=1$, Fig~\ref{fig:flow_curves}b, and for an onset stress that occurs at very high St, Fig~\ref{fig:flow_curves}c. In each case, a transition is observed between the frictionless and frictional states, similarly to the totally viscous ($\text{St} \ll 0$) case. In Fig~\ref{fig:flow_curves}b, the frictionless regime is observed for Stokes numbers up to around unity. Below this point, the suspension viscosity is independent of the Stokes number. For larger Stokes numbers, we observe frictional, inertial flow, with $\sigma_{xy}/\eta_f \dot{\gamma} \propto \dot{\gamma}$. The linear scaling of viscosity with shear rate above St$=1$ is due to inertial effects; the larger jump in viscosity (i.e. the super-linear behaviour between $\dot{\gamma}/\dot{\gamma}_0=0.1$ and $\dot{\gamma}/\dot{\gamma}_0=1$) is due to the onset of frictional contacts. We verify that the flow in each of these limits remains frictionless and frictional respectively at the microscale by examining the fraction of particle contacts that have exceeded $F^{CL}$. It is noted that the result in Fig~\ref{fig:flow_curves}b corresponds directly to the shear thickening phenomenon observed in the simulations reported by Fernandez\cite{Fernandez2013}, with a low shear rate regime in which lubrication dominates and particle friction is absent and a high shear regime dominated by friction with a viscosity that depends linearly on the shear rate. Fernandez also reports experimental findings for shear flow of polymer coated quartz miroparticle suspensions that appear qualitatively similar to Fig~\ref{fig:flow_curves}b, though it is not clear whether the Stokes number is appropriate for such a comparison. Indeed, a similar set of experimental findings~\cite{Melrose1996} were previously attributed to enhanced \emph{dynamic} friction due to increased resistance to fluid flow in the polymer layer.
In Fig~\ref{fig:flow_curves}c, the onset stress occurs at St $\approx300$, so both the frictionless and frictional states (again, these are verified by appealing to the frictional of individual contacts) exhibit $\sigma_{xy}/\eta_f \dot{\gamma} \propto \dot{\gamma}$ scaling, with super-linear behaviour representing the frictional transition. In each case, we obtain from the simulations excellent agreement with theoretical predictions in the limits of fully frictionless and fully frictional flow.


These novel flow curves clearly illustrate the distinction between shear thickening driven by friction and by inertia. By tuning the particle and fluid properties appropriately, we have demonstrated that although (experimentally) the regimes may seem highly distinct and therefore challenging to achieve within a single system, both mechanisms might be made to arise concurrently, giving rise to new rheological behaviour. The challenge remains to achieve a sufficient understanding of the roles of particle surface chemistry, particle size and suspending fluid properties to realise and utilise these flow regimes experimentally.

\begin{figure}
      \subfigure[]{\includegraphics[width=75mm]{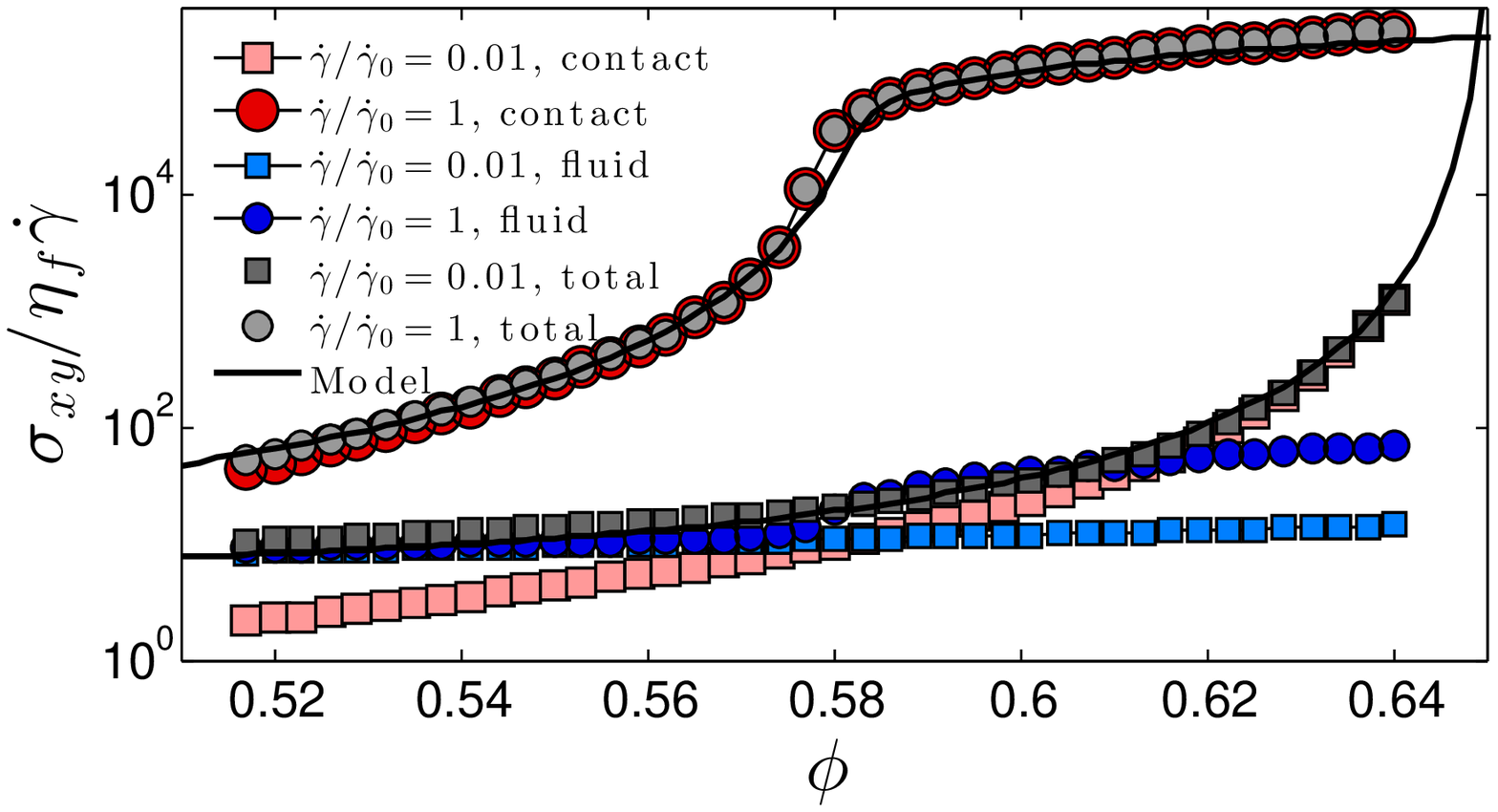}}
      \subfigure[]{\includegraphics[width=75mm]{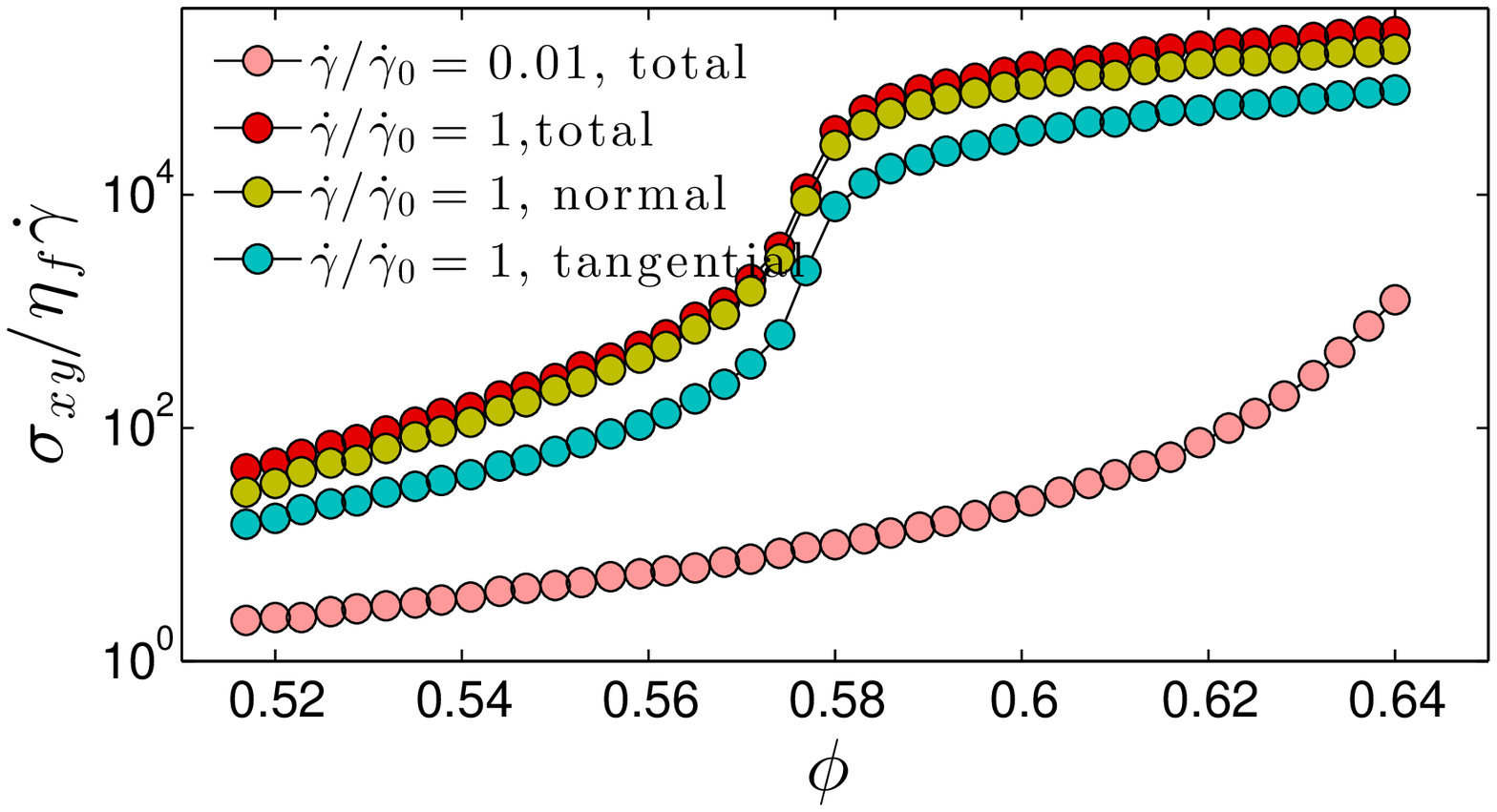}}
     \caption{
(a) Divergence of viscosity contributions and model prediction;
(b) Divergences of normal and tangential contact forces.
 }
 \label{fig:divergence}
\end{figure}

For the entirely viscous case presented in Fig~\ref{fig:flow_curves}a, we isolate the contact and fluid contributions to the viscosity and plot them against volume fraction for the frictionless ($\dot{\gamma}/\dot{\gamma}_0 = 0.01$) and frictional ($\dot{\gamma}/\dot{\gamma}_0 = 1$) limits in Fig~\ref{fig:divergence}a. It is noted that analogous results are obtained for the inertia-dominated cases, though the magnitude of the viscosity is increased consistent with the linear viscosity scalings demonstrated in Figs~\ref{fig:flow_curves}b-c. Comparing the jumps from the frictionless to the frictional branch,  it is demonstrated that the main increase in viscosity upon shear thickening is due to the contact contribution, while there is only a very minor increase in the fluid contribution (appreciable only at high volume fractions). While this suggests a configurational change leading to a change in the mean fluid film thickness or film number (resulting in a slightly increased fluid stress), it is not consistent with the notion of a large macroscopic transition to hydroclustering~\cite{Wagner2009} and a corresponding massive increase in viscous dissipation. We further decompose the contact stress into normal and tangential components, Fig~\ref{fig:divergence}b. We find that although the major difference between the frictionless and frictional limits at the individual particle level is the presence of tangential contact forces, the main contributor to the increase in the contact stress is in fact the normal component, rather than the tangential component, further corroborating the major role played by the particle configuration change induced by friction. This change can be understood from the perspective that the available degrees of freedom for particle motion decrease at the onset of frictional contacts, in that frictional particle assemblies require four contacts per particle for mechanical stability, while frictionless ones require six~\cite{Song2008}. At fixed volume fraction, the transition to frictional behaviour is therefore manifested as an increased resistance to flow that necessitates greater particle overlaps and results in higher particle pressure. Interestingly, reducing the available degrees of freedom by means other than particle friction leads to the same observation. In a separate simulation we model steady shear at $\dot{\gamma}/\dot{\gamma}_0 = 0.01$, and at time $t_1$ we set all $z-$components of particle velocity and forces to zero, effectively imposing a 2D flow constraint. A large increase in the particle contribution to the stress is observed, consistent with the shear thickening behaviour presented here, that dissipates at a later time $t_2$ when the 2D flow constraint is relaxed. The dominant role of contacts, and the sensitivity to their nature (whether tangential forces may be sustained in addition to normal forces), remains a contentious issue; these results add further weight to the argument for frictional contacts as a crucial contributor to (non-inertial) shear thickening.

We have therefore demonstrated that the frictional thickening mechanism, modelled via a load-dependent particle friction in DEM and captured by the constitutive model, can occur within a variety of flow regimes and may thus couple or compete with inertial thickening. Both the DEM simulation and the constitutive model capture consistently such shear thickening phenomena. By isolating the contact and hydrodynamic contributions to the shear stress, we have shown that the shear thickening transition is heavily dominated by particle contacts as opposed to hydrodynamic effects. The large increase of contact stress upon shear thickening has been attributed mainly to normal contact forces, though we note that the tangential contact forces present in the thickened regime largely exceed the normal forces present in the non-thickened state.

\subsection{Microscopic analysis}
\label{sec:micro}

We use microscale information to further characterise the distinction between the thickened (frictional) and non-thickened (frictionless) states. The microstructure is quantified using the particle-particle contact number and the fabric, or net-orientation, of these contacts, while particle-level dynamics are quantified using correlation functions in displacement and velocity.
Work in rheo-imaging of colloidal systems~\cite{Isa2007,Ballesta2008} has demonstrated the potential for these dynamic properties to be obtained and quantified experimentally for shear thickening materials. In addition, further experiments~\cite{Clara-Rahola2012a} have led to their successful quantification for a sheared, highly polydisperse emulsion, using confocal imaging. We anticipate that future such analyses for shear thickening suspensions will benefit  from, and corroborate, the results presented in this paper. 

\begin{figure*}
\centering
  \includegraphics[width=88mm]{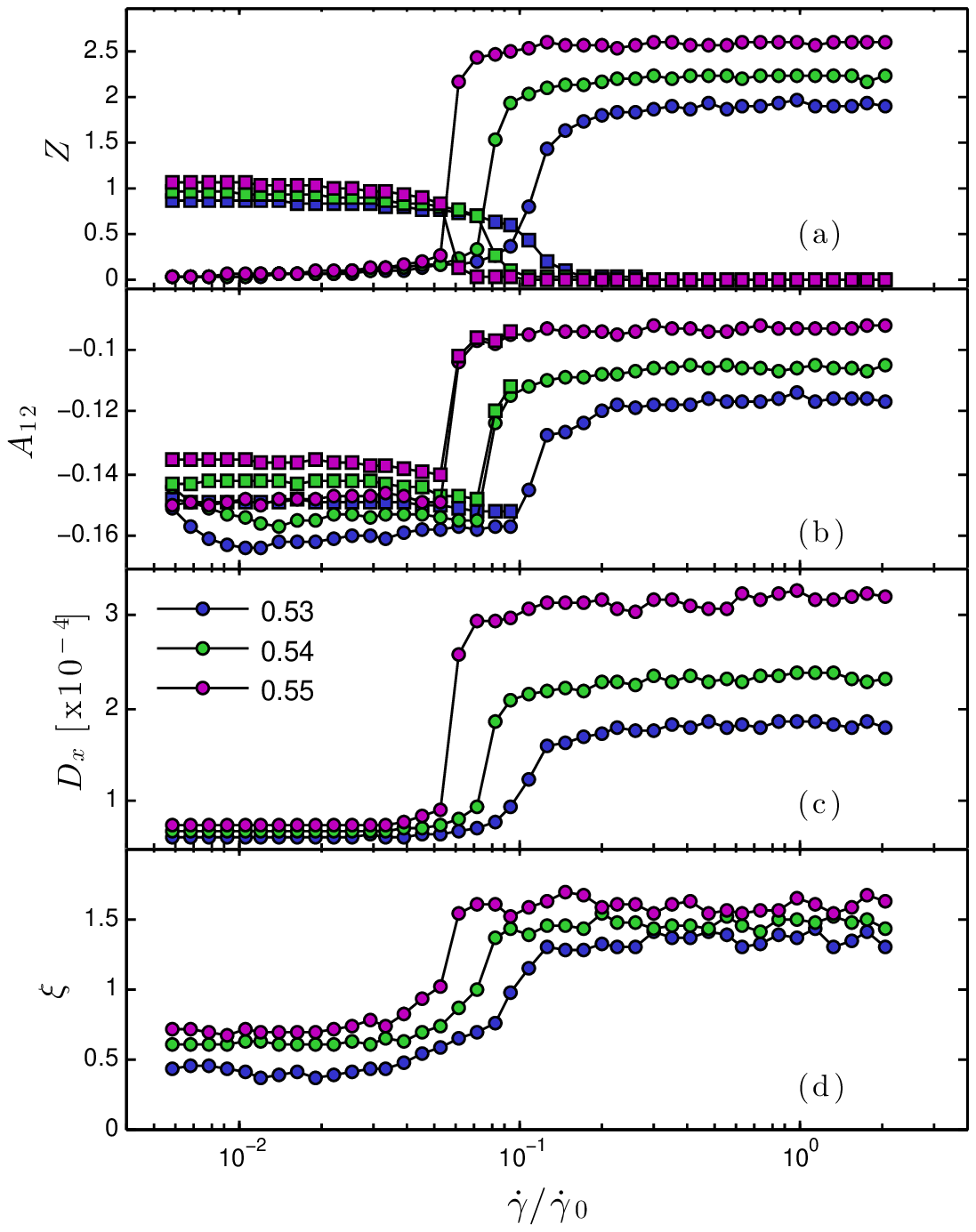}
  \includegraphics[width=88mm]{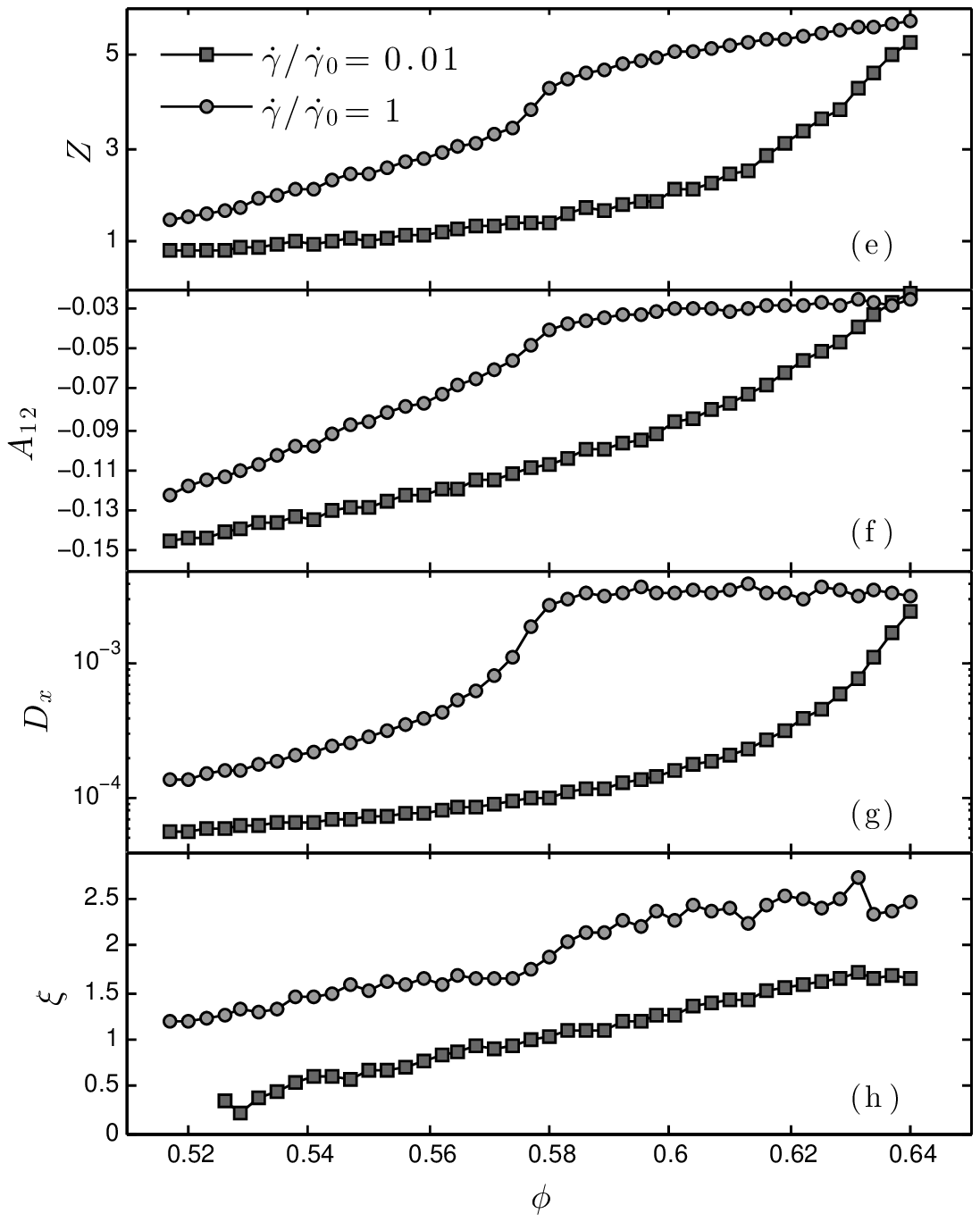}
     \caption{
(a)-(d) Evolution of microscale structures and dynamics across the frictional shear-thickening transition.
(a) Mean number of particle-particle contacts. Squares represent low force contacts for which friction is not activated; circles represent high force contacts for which friction is activated.
(b) Shear component of the fabric tensor. Squares represent low force contacts for which friction is not activated; circles represent high force contacts for which friction is activated.
(c) Effective diffusion coefficient.
(d) Velocity correlation length as defined by Eq~\ref{eq:correlation_length}.
(e)-(h) Evolution of microscale structures and dynamics with volume fraction, for frictional and non-frictional states.
(e) Mean number of particle-particle contacts. Stars represent low force contacts for which friction is not activated; circles represent high force contacts for which friction is activated.
(f) Shear component of the fabric tensor. Stars represent low force contacts for which friction is not activated; circles represent high force contacts for which friction is activated.
(g) Effective diffusion coefficient.
(h) Velocity correlation length.
 }
 \label{fig:microstructure_a}
\end{figure*}

The microstructure is characterised based on two separate length scales. For the contact number $Z$, we calculate (a) the average number of frictionless, $|\mathbf{F}^{c,n}_{i,j}|<F^{CL}$ interactions per particle; (b) the average number of frictional, $|\mathbf{F}^{c,n}_{i,j}|>F^{CL}$ contacts per particle. For the contact fabric, we adopt the formulation used by Sun and Sundaresan~\cite{Sun2011}
\begin{equation}
\mathbf{A} = \frac{1}{Z} \sum^{\text{N}_c}_{\alpha=1} \mathbf{n}_\text{ij} \mathbf{n}_\text{ij} - \frac{1}{3} \mathbf{I} \text{.}
\end{equation}
Under shear flow, contacts preferentially align along the compressive axis at (or close to) $45^{\circ}$ to the $x-$ and $y-$axes (the flow and gradient directions respectively), with the corresponding shear component of $\mathbf{A}$, $|A_{12}|$ quantifying the extent of the anisotropic alignment. $A_{12}=0.5$ represents perfect alignment of all contacts in the compressive axis, while $A_{12} = 0$ represents perfect isotropy. As with the contact number, we quantify the shear component of the fabric based on both the network of frictionless contacts and that of frictional contacts.
The non-affine motion is quantified by first obtaining a coarse-grained velocity profile for the shearing flow and subtracting the appropriate value of this coarse-grained velocity from each particle's velocity vector to obtain the ``fluctuating'' velocity component. From these fluctuating velocities we obtain the mean squared displacement (MSD), averaged across particles and time steps. Plotting the MSD ($\left<{x}^2\right>$) versus strain magnitude ($\dot{\gamma}t$) yields a linear diffusive behaviour that is independent of Stokes number, for the case of zero inertia. We take the gradient of this slope $\text{D}_x$ as an \textit{effective} diffusion coefficient ($i.e.$ $\left<x^2\right>=\text{D}_x \dot{\gamma}t$), though strictly the diffusion coefficient in this case is $\text{D}_x \dot{\gamma}$. 
In addition, we calculate the correlation of the fluctuating velocity vectors according to Lois~\cite{Lois2007a,Sun2006,Ness2015}
\begin{equation}
c({r}) = \frac{\sum_i \sum_{j>i} \overline{\mathbf{v}}_i \cdot \overline{\mathbf{v}}_j \delta(|\mathbf{r}_{ij}| - {r})}{\sum_i \sum_{j>i}\delta(|\mathbf{r}_{ij}| - {r})} \text{,}
\label{eq:correlation_length}
\end{equation}
where $\overline{\mathbf{v}}_i$, $\overline{\mathbf{v}}_j$ are particle velocity vectors averaged over a length of time sufficient to give an averaged particle displacement due to the mean flow of approximately $0.5d$. It is found that the correlation decays approximately exponentially with the distance between particle centres $r$. We therefore fit a simple functional form $C(r) = ke^{-r/\xi}$, where $\xi$ takes units of particle diameter and is hereafter referred to as the ``correlation length'', and $k$ is a constant prefactor.
The evolution of these microscale quantities is presented in Fig~\ref{fig:microstructure_a} as a function of $\dot{\gamma}/\dot{\gamma}_0$ at three volume fractions, and as a function of volume fraction at $\dot{\gamma}/\dot{\gamma}_0 = 0.01$ (frictionless) and $\dot{\gamma}/\dot{\gamma}_0 = 1$ (frictional), for the case of zero inertia.

In Fig~\ref{fig:microstructure_a}a we demonstrate the increasing number of frictional contacts and the diminishing of frictionless interactions as the shear rate is increased and the onset stress is exceeded. The results presented here are consistent with the evolution of the fraction of frictional contacts presented by Mari et al.~\cite{Mari2014a}. It is further noted that for a fixed volume fraction, the number of direct particle-particle contacts that exist in the frictional, shear thickened state exceeds the number of frictionless, normal interactions that were present in the non-thickened state. This suggests that in the process of becoming frictional, the particles have rearranged into a distinct microstructural configuration.
The evolution of $A_{12}$, Fig~\ref{fig:microstructure_a}b, confirms this. In the non-thickened state, we observe distinct microstructures for the networks of frictionless and frictional contacts, though the number of frictional contacts is very small. At each volume fraction, contacts for which $F^{CL}$ is exceeded tend to be aligned more strongly with the compressive flow direction than those contacts for which friction is not activated. Upon shear thickening, however, the fabric of the frictional contact network moves closer to zero, while the frictionless fabric disappears due to an absence of such interactions far above the onset stress. 
The microstructural information suggests that the shear thickening transition brings the particle configuration closer to what might be expected for a quasistatic, rate-independent flow, where $Z \approx 4$ and $A_{12}\approx -0.03$~\cite{Sun2011,Song2008} for the frictional cases. Shear thickening can therefore, in this sense, be considered analogous to an increase in volume fraction at constant friction, in that the central change in each case is that the departure from the critical volume fraction, quantified as $|\phi - \phi_c|$, decreases. It is noted that the inertial cases exhibit very similar behaviour with respect to the microstructural properties. The exception is a very modest increase in contact number, smaller than 10\%, for the inertia-dominated flows.

Particle level dynamics are found to exhibit analogous behaviour across the thickening transition. In Fig~\ref{fig:microstructure_a}c, we show that for each volume fraction studied, there is a significant jump in the effective diffusion as the suspension shear thickens. As with the microstructure, this is consistent with the suspension becoming closer to its jamming volume fraction as the shear rate or stress is increased. As the extent of frictional behaviour increases, particles form more contacts and are required to deviate further from an affine trajectory in order to pass each other as they are subjected to shear at the same or higher rate. The average displacement deviation from the mean flow therefore undergoes a step change. A similar transition is observed in the velocity correlation length, Fig~\ref{fig:microstructure_a}d, which indicates that in the shear thickened regime, particle trajectories tend to be more correlated with those of their immediate neighbours, suggesting a tendency towards collective motion of particle groups. Though this is, indeed, qualitatively reminiscent is some respects to the ``hydroclustering'' behaviour previously reported~\cite{Wagner2009}, we note that the particle groups that collectively move in the present simulations are found to be unanimously under frictional contact, rather than separated by lubrication layers. We therefore suggest that while clustering is apparent, it is, in this case, more accurately described as \textit{frictional-} rather than \textit{hydro-}clustering. Indeed, experimental techniques that purportedly demonstrate hydroclustering are in fact unable to distinguish between these two mechanisms of dynamic correlation~\cite{Cheng2011}.

We demonstrate in Fig~\ref{fig:microstructure_a}b that each of the transitions presented in Fig~\ref{fig:microstructure_a}a corresponds to a shift between distinct and well-defined branches in each of the microscale parameters investigated, in the thickened ($\dot{\gamma}/\dot{\gamma}_0 = 1$) and non-thickened ($\dot{\gamma}/\dot{\gamma}_0 = 0.01$) limits. In addition, it is observed that the differing divergences of each pair of branches with volume fraction is consistent with that observed for the bulk suspension viscosity, namely the frictional, high stress branch diverges near $\phi = \phi_m$ and is thereafter consistent with quasistatic behavior~\cite{Sun2011}, while the frictionless, low stress branch diverges towards $\phi = \phi_{RCP}$. We note that the divergence is less clear for the correlation length $\xi$ than for the other microscale parameters investigated, which can be attributed to the relatively small domain size, which places limits on the length scale over which correlations can be observed. The distinction between branches, however, is convincing.
A demonstration of such contrasting microscale divergences in the thickened and non-thickened rheological regimes in an experimental system would prove invaluable in corroborating this work, and in highlighting the essential role of friction as the origin of the distinct rheologies below and above shear thickening.


\section{Concluding remarks}
In this paper we have explicitly demonstrated the distinction between frictional and inertial shear thickening mechanisms, and illustrated their presence as separate regimes on the broader flow map of dense suspensions. In practice, frictional shear thickening is typically observed for colloidal ($d \lesssim 1 \  \mathrm{\mu m}$) suspensions for which inertia is always absent (or negligible), while the inertial shear thickening typically occurs in granular $d \gtrsim100 \  \mathrm{\mu m}$) suspensions for which friction is always present (or more accurately starting from exceedingly low Stokes numbers). Our simulation results suggest that in principle thickening may occur in a mixed mode with both mechanisms playing a role. This may indeed be the case for a suspension of mixed colloidal and granular particles, as hinted by the recent experiments on shear thickening with intermediate particle sizes~\cite{Guy2015a}, which indicate that the frictional thickening onset stress scales with the inverse square of particle size. There are of course many other possible scenarios where mixed thickening could occur as suggested from our simulations, though it remains to be seen whether such rheology can be observed experimentally.

Transitions in the microstructural and dynamic variables are observed across the frictional thickening transition, and we have shown that the microstructure of the shear thickened and non-thickened states exhibit distinct divergences with respect to volume fraction,  indicating the microscale mechanism for the same behaviour of the bulk suspension viscosity. We expect the results presented here to provide useful means of analysing new results obtained from particle microscopy and imaging of model experimental systems.


\section*{Acknowledgements}

This work is funded by the Engineering and Physical Sciences Research Council (EPSRC) and Johnson Matthey through a CASE studentship award. The authors would like to thank M. Marigo, P. McGuire, H. Stitt, H. Xu, J.Y. Ooi, B. Guy, M. Hermes, W.C.K. Poon and M.E. Cates for helpful discussions.

\bibliography{../../../../../00_references_master/library} 

\clearpage
\appendix
\onecolumngrid

\end{document}